\documentclass[conference]{IEEEtran}
\IEEEoverridecommandlockouts
\usepackage{cite}
\usepackage{amsmath,amssymb,amsfonts}
\usepackage{algorithmic}
\usepackage{graphicx}
\usepackage{textcomp}
\usepackage{xcolor}
\usepackage{lipsum}
\usepackage{float}
\usepackage{multirow}
\usepackage[breaklinks=true]{hyperref}
\usepackage{breakcites}
\usepackage[right=0.62in, left=0.62in, top=0.75in, bottom=1.0in]{geometry}

\def\BibTeX{{\rm B\kern-.05em{\sc i\kern-.025em b}\kern-.08em
    T\kern-.1667em\lower.7ex\hbox{E}\kern-.125emX}}
    
\begin{document}
    \bstctlcite{IEEEexample:BSTcontrol}
    \title{Intelligent Routing as a Service (iRaaS)\\
    \large A Flexible Routing Framework for Knowledge-Defined Networks
    \thanks{This work is a contribution by Project REASON, a UK Government funded project under the Future Open Networks Research Challenge (FONRC) sponsored by the Department of Science Innovation and Technology (DSIT) \\ 
    ISBN 978-3-903176-63-8 © 2024 IFIP}
    }
    \author{\IEEEauthorblockN{Saptarshi Ghosh \IEEEauthorrefmark{1},Konstantinos Antonakoglou \IEEEauthorrefmark{1}, 
    Ioannis Mavromatis\IEEEauthorrefmark{1}, and Kostas Katsaros \IEEEauthorrefmark{1}\\
    \IEEEauthorblockA{\IEEEauthorrefmark{1} Digital Catapult, United Kingdom\\}
    Emails: \{saptarshi.ghosh, konstantinos.antonakoglou, ioannis.mavromatis, kostas.katsaros\}@digicatapult.org.uk}}
    
    \maketitle
    
    \begin{abstract}
        The scope of the Sixth-Generation Self-Organized Networks (6G-SON) advances its predecessor’s capability towards agility, flexibility, and adaptability. On-demand overlay networking technologies have shown a prominent maturity while coping with the rising complexity and scale of enterprise, service provider, and data centre networks. The Software-Defined Networking paradigm has recently offered Model Driven Programmability, minimizing network management complexity through automation and orchestration. However, leveraging Machine Learning-driven network optimization, a.k.a. Knowledge-Defined Networking (KDN), has still been a domain of interest for the Network Softwarization research community. 
        In this article, we propose Intelligent Routing as a Service (iRaaS) architecture as an application layer cognitive routing framework for KDNs. iRaaS offers routing logic customization (i.e., customizing metric function and path-finding algorithm)  and provides an option to include heuristic parameters from trained models as a part of the metric calculation. iRaaS sits on the application plane above the knowledge plane in a KDN stack, thus providing platform- and vendor-agnostic coupling with existing network infrastructures. This article covers the scope of iRaaS by using reliability as a heuristic for standard path-discovery algorithms, e.g., Shortest Path First (SPF) and Diffusion Update algorithm (DUAL), along with the architectural specification. We validate our approach through a Proof-of-Concept deployment. 
    \end{abstract}
    
    
    \section{Introduction}
        As computer networks become increasingly complex, manual network optimisations become less feasible. As a result, many organisations have turned to network automation, which offers improved efficiency and reduced human error. Network automation configures, provides, manages, and tests devices and systems, improves effectiveness and redundancy, and meets compliance standards. With the active rollout of the 5G ecosystem, Softwarization, Virtualization, Cloudification (RFC 7868), and Interoperability have been the prominent adaptations. The 5G Infrastructure Public Private Partnership (5G PPP) architectural working group identifies two new stakeholders in the domain, namely, Virtualization Infrastructure Service Provider (VISP) and Data-Centre Service Provider (DCSP). Cloud-native architecture, a de-facto standard for contemporary Network Functions Virtualization (NFV) deployment, offers agile, lightweight, and manageable solutions. It seamlessly blends DevOps principles and practices, providing a flexible and scalable environment for network operations. As a result, consumer enterprises opt for features like Zero Touch Provisioning (ZTP), which offloads a significant amount of network administration burden to the service provider end, where the networks are centrally managed and orchestrated with predefined policies. To accommodate this transition, centralised optimisation algorithms (e.g., Routing, Quality-of-Service (QoS)), infrastructure automation tools (e.g., Ansible, Puppet, Chef), high-availability protocols (e.g., first-hop redundancy protocol (FHRP), Stateful Switchover (SSO)), and ML-based network-state prediction models (e.g., Time-Series analysis, Traffic-classification, route-prediction) play a crucial role. 

As mentioned above, the developments demand that the underlying telecommunication infrastructure be flexible and self-aware. Although proposed in IMT-2020, the Ultra-Reliable and Low Latency Communications (URLLC) verticals have yet to be achieved entirely. Due to spectrum limitations, further suppression of the data plane latency has become challenging. Therefore, research has turned to optimising the control plane using Routing Optimization, which aims to reduce latency due to the optimal path-finding process. Nevertheless, RO enablers such as rapid converging routing protocols, optimal route-redistribution, route reliability estimation, and Link-State prediction bring forth scalability and flexibility challenges when considering deploying multi-vendor, multi-protocol, elastic, and dynamic networks. Historically, routing protocols (e.g. OSPF, EIGRP, ISIS) do not support centralised computing models, and commercially available routing protocols that natively support a Software Define Network (SDN) and Knowledge Defined Networking (KDN) model are limited if the present context is concerned.

In summary, as network automation and orchestration thrive to fulfil the scale of deployment required to cope with service quality demand, a centralised routing model with programmable routing logic to offer service-specific customisation shows excellent potential in optimising the quality of experience by offering Routing-as-a-Service (RaaS). 
        
   In this paper, we address the aforementioned reliability, scalability and flexibility challenges of such cases of Self-Organized Networks (SON). We present an intent-based, data plane-agnostic and intelligent Routing-as-a-Service platform that:
        \begin{itemize}
            \item Allows the deployment of customized routing logic and enables the life-cycle management and use of ML models for optimizing network reliability.
            \item Provides a centralized service-based architecture overseeing the entire network, diminishing the time-consuming control plane packet exchange, hence reducing control plane-induced latency. 
            \item Captures the declarative requirements of the platform user and the end-to-end network topology abstracted by the underlying SDN/KDN and legacy controller-less infrastructure.
            \item Uses robust telemetry to capture the state of the end-to-end network.
            \item Adopts state-of-the-art standards and open-source solutions, validating our solution's sustainability and extensibility. 
        \end{itemize}
       
        The remainder of this paper is the following, section II gives a background of the context, section III describes the high-level architecture of iRaaS, section IV provides details of the iRaaS system design with sequence diagram and a bespoke telemetry architecture named ShellMon, section V validates the iRaaS architecture with a proof of concept testbed setup related results, Finally, Section VI concludes this article with a summary and future scope aimed for this work.

    \section{Background}
    \label{Section_Background}
This section will explore the bottlenecks of distributed computing models utilised in traditional routing protocols, particularly regarding rapid convergence and scalability. Subsequently, with a few contemporary examples, it delves into how the SDN paradigm offers a framework to implement centralised routing as a possible solution. Finally, it highlights the advantages of application layer routing as a basis for RaaS.

\subsection{Limitation of traditional IP Routing Protocols}
Dynamic networks, such as Mobile Ad-hoc Networks (MANET), categorize routing protocols into three main types: proactive, reactive, and hybrid. These protocols update the routing table regularly or when the network topology changes. Traditional IP routing protocols have limited flexibility in customizing the metric formulation; for instance, they do not consider radio link characteristics like SNIR and path-loss in route calculation. To optimize IP routing over diverse radio networks and provide users with real-time access to critical information on the move, Cisco introduced the Radio Aware Routing (RAR) \cite{soa_fn_3}. The latest RAR protocol, Dynamic Link Exchange Protocol (DLEP), has been standardized in IETF (RFC 8175). Cisco has also worked towards routing optimization in MANET \cite{soa_fn_4}. Furthermore, OSPFv3 enhances routing efficiency and reduces overhead traffic in MANET environments so that network clusters can scale to support more users.

The Enhanced Interior Gateway Routing Protocol (EIGRP) uses a hybrid approach with the Diffusion Update Algorithm (DUAL) to combine distance vector and link-state routing mechanisms. This approach helps to achieve rapid convergence by pre-calculating a subset of neighbours called Feasible Successors (FS) as the next hop for each destination prefix. EIGRP also offers a more comprehensive composite metric function that considers effective throughput, latency, load, reliability, and jitter, which is superior to OSPF, which only considers bandwidth. However, EIGRP may only provide rapid convergence if neighbours qualify as an FS or if the FS becomes available. In such cases, EIGRP suspends its data plane until it finds a loop-free alternate path by recursively querying its neighbours. The dynamic metric attributes load and reliability are also not used in practice as they may cause route-flapping.

As computer networks grow larger and more complex, traditional hardware-based routing may result in slower convergence and decreased network performance. This is because routing protocols rely on distributed algorithms that require communication systems to exchange control information, which can increase overhead and become a bottleneck for maintaining efficiency at scale.

\subsection{The SDN Paradigm for Centralised Routing}
The Software-Defined Networking (SDN) paradigm addresses the limitation mentioned above by decoupling the control (CP) and Data Plane  \cite{rfc_sdn}. In SDN architecture, a logically centralised CP oversees a distributed DP, which provides a platform for centralised routing models to operate, diminishing the overhead of control packet exchange at the DP. Moreover, the central placement of the SDN controller simplifies the CP topology from mest to tree. Consequently, The SDN controller exhibits the potential to leverage the link-state routing approach to guarantee loop-free paths in a space-efficient centralised computing model with no control overhead at the DP. 

In an SDN, forwarding devices like routers and switches at the DP report their local neighbourship information to the controller using the south-bound interface (SBI) without hopping through any neighbours. This minimises the communication complexity of the control plane invariant to the network diameter. The controller builds a graph locally representing the underlying network topology using link-state logic from the reported neighbourship information. Therefore, path calculation algorithms can compute the routes without any control packet exchange, resulting in faster convergence. Thus, forwarding devices can request the controller for path discovery, and with pre-computed paths, the controller can respond by meeting a hard deadline. Therefore, a stable SBI (i.e., with negligible jitter) and a proactive controller (i.e., with pre-computed routes) can provide rapid convergence with a hard deadline, an essential requirement for URLLC. 

From the perspective of Internet routing, benefits of Routing-as-a-Service have been presented in \cite{Lakshminarayanan:CSD-04-1327} explaining how it can resolve the conflict of path selection for satisfying the QoS requirements of end-to-end network users while allowing Autonomous System administrators (e.g. ISPs) to control traffic flows over their entire infrastructure.

Incorporating SDN, a RaaS platform in \cite{BU2019130} allows the platform user to select routing algorithms used as network functions to compose a customised routing service. Nevertheless, this solution relies on existing routing protocols.

A Software-defined Wide Area Network (SD-WAN) platform, presented in \cite{9700661}, calculates optimal paths using CPLEX over Open Network Operating System (ONOS)-controlled routers with possible extension to ML-based optimal path calculation. Nevertheless, it does not provide a way to customise the metric function for calculating the optimal path and lacks a telemetry architecture that any potential ML model will require.

\subsection{Application Layer Routing}
        Application layer routing is particularly beneficial for heterogeneous SDNs, where the network infrastructure is diverse and programmatically controlled. In such environments, application layer routing leverages the centralized intelligence of SDN controllers to make dynamic, application-aware decisions. This approach enables the network to adapt in real-time to changing application demands and network conditions, optimizing for factors like bandwidth, latency, and security requirements specific to each application. For heterogeneous SDNs, which may span across different \textit{domains} and incorporate a variety of physical and virtual network functions, application layer routing ensures that traffic is efficiently and intelligently routed, taking advantage of the programmable nature of SDNs to enhance performance, scalability, and resilience. By aligning network behaviour with application requirements, it facilitates a more responsive and optimized network environment tailor-made for the diverse needs of modern digital applications.

    \section{High-level System Architecture}
    \label{sec:architecture}
     \begin{figure*}[ht]
            \centering\includegraphics[keepaspectratio=true,scale=0.24]{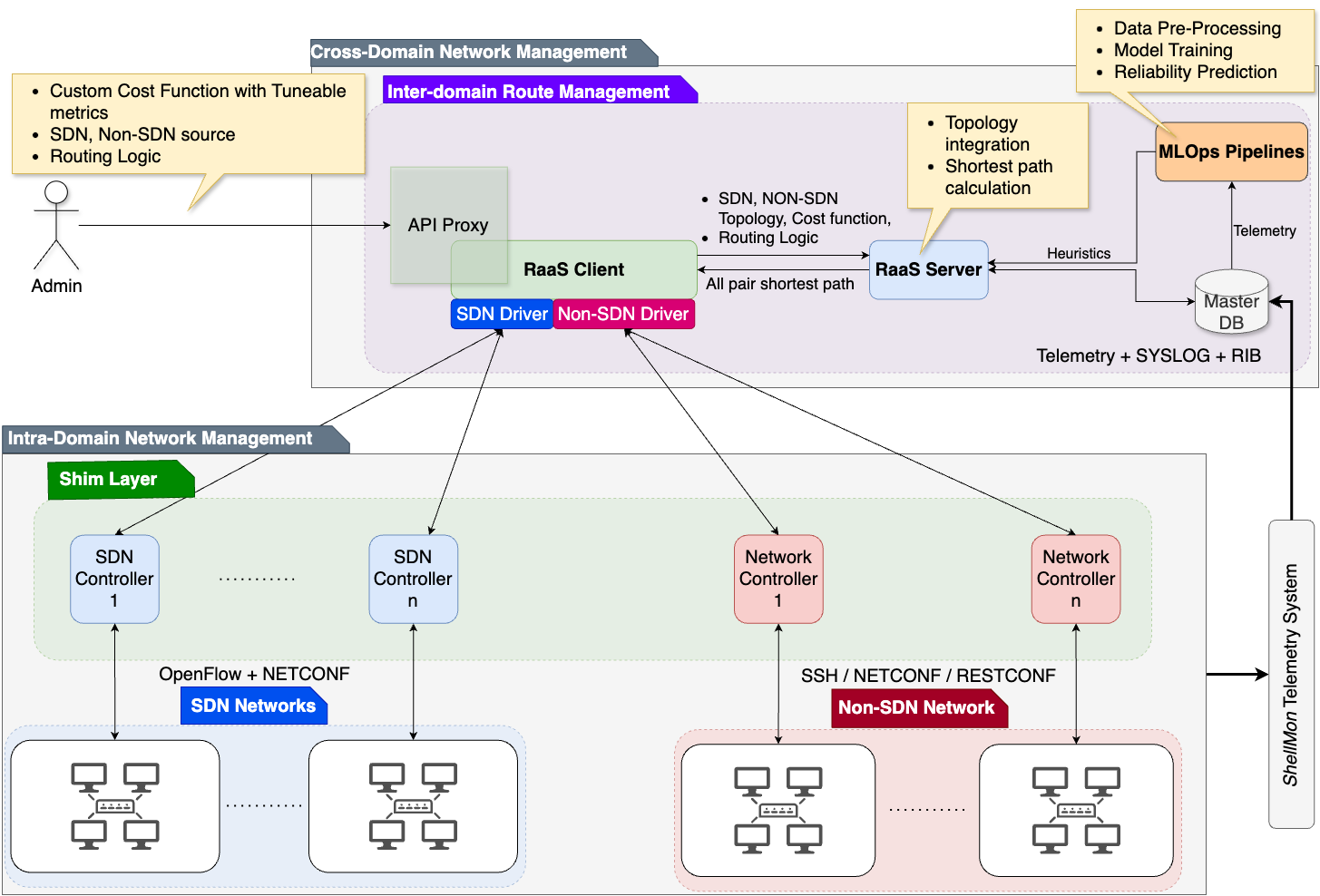}
            \caption{The proposed iRaaS System Architecture. The system consists of multiple domains and a cross-domain management plane that is responsible for the orchestration and routing across each domain.}
            \label{fig:RaaS_SA}
        \end{figure*}

        Based on the analysis in the previous section, we envision a system that extends the standard SDN implementation and provides routing in an "as-a-service" fashion. This implies that the routing service will be provided as an application remotely accessible by end-users and administrators and deployed in a softwarized fashion. 
        Starting by the definition of the \textit{domain}, we imply a system which is a "component subsystem'' of a more comprehensive system where the exposure of internal design and operations of the system external to the domain are limited. This could, for example, be an administrative domain (e.g., an Internet Service Provider (ISP)). We later envision that there will be a management plane responsible for the intelligent intent-based routing across the different \textit{domains}. This architecture can be seen in Figure \ref{fig:RaaS_SA}. This figure also introduced our key functional blocks that provided the envisioned capability described in this paper. As per our envisaged functionality, we propose implementing a centralized application-layer routing model that accompanies and extends traditional SDN and non-SDN controllers, enhancing the system's performance, scalability, and resilience.
        Our envisaged iRaaS application sits within a Cross-Domain Network Manager (CDNM) responsible for orchestrating the different services deployed there. The Cross-Domain Route Management (CDRM) takes the routing decisions and communicates them downstream to the administrative domain controlled by an Intra-Domain Network Manager (IDNM). The domain comprises a hybrid SDN network spanning multiple controllers (i.e., SDN and Non-SDN). We refer to the combined control plane of SDN and Non-SDN controllers within an administrative domain as Shim-Layer, as it abstracts the platform specificity of the underlying data plane from the planes above. We break down our application layer into two entities, i.e., the iRaaS Client and the iRaaS Server, to enhance load balancing and enable higher scalability. Moreover, we propose a telemetry application that collects real-time KPIs from the different domains and sends them to the RaaS server to enable intelligent decisions. The following sections describe the system in detail.
        
        \subsection{iRaaS Client}
            iRaaS client receives Route Intent from external admins through an API proxy, which enables an administrator (human or program) to interact through a single point of contact and select common routing attributes, e.g., routing protocol and its associated parameters and path manipulation logic. RaaS offers an admin to customize routing logic. A custom routing logic can be known as the Shortest Path Algorithm (e.g., SPF, DBF and DUAL) or a bespoke one encoded in compliance with the iRaaS Server API. Additionally, it offers the ability to customize the cost function (e.g., an admin might use SPF as routing logic for implementing LSR in a hierarchical topology with EIGRP-like composite metric). This level of flexibility of iRaaS contributes to its novelty and uniqueness. The iRaaS client is also responsible for building an aggregated graph of the underlying topologies of SDN and non-SDN available controllers. The iRaaS Client sends the Route Intent and aggregated graph to the iRaaS Server.

            The iRaaS architecture supports hybrid SDN at the access plane. The admin informs the iRaaS client about the respective controllers’ management interface address while requesting through the API Proxy. The client establishes management access with all controllers at the access network leveraging the CaaS service and respective drivers (SDN/Non-SDN) and fetches the controller-wise downstream topologies through standard management protocol, e.g., NETCONF \cite{rfc6241}, and RESTCONF \cite{restconf}). Finally, the northbound interfaces between the iRaaS Client and the Shim layer may use an exterior gateway transport for connectivity. 
        
        \subsection{iRaaS Server}
            The iRaaS server receives the Route Intents (i.e., shortest path algorithm and cost function) and a graph representing the underlying topology from the iRaaS client through a standardised interface. The server then weighs the graphs using the link cost reported from the telemetry system using the cost function defined at the intent. The iRaaS Server then calculates the optimal path(s) following the routing logic. Predictive analytics, such as reliability estimation, uses an MLOps pipeline, which contributes to cost calculation. The iRaaS server then responds to the iRaaS Client with a set of paths, which the client informs the downstream controllers to install at the forwarding devices. The MLOps pipeline automates the standard machine learning workflow of data preprocessing, training, testing and validation. The paper \cite{kdn} shows a Sharpe Ratio-based path-reliability estimation using a Recurrent Neural Network (RNN) with Long Short-Term Term Memory (LSTM).
        
        \subsection{Telemetry Framework}
        The telemetry framework resides in conjunction with the administrative domains and the IDNM and provides a platform and vendor-agnostic multi-modal implementation that collects data in a standardized fashion. The telemetry framework exposes a number of RESTful interfaces and publish-subscribe messaging buses responsible for collecting telemetry data and storing them in the database available in the CDRM. In the following section, we go into more detail on the implementation of the telemetry framework describing the different communication modes supported. 

        \subsection{Interfacing with 5G Systems}
        Although this paper does not demonstrate interfacing with a 5G system. That said, the iRaaS framework can influence packet routing at the 5G user plane. In such scenarios, the iRaaS client would communicate with the Network Exposure Function (NEF) to perform Policy-Based Routing (PBR) by modifying routing tables held at the User-Plane Function (UPF). To achieve rapid convergence, NEF can order the redundant paths received from the iRaaS server by their cost and install them in the UPF routing table as floating static routes.
        
        \section{System Design and Provided Functionality}
        Following the above high-level architecture, in this section, we describe in more detail how iRaaS is provided by our system and describe all function blocks that build up our platform. As briefly discussed earlier, our proposition supports both traditional SDN implementations and monolithic implementations that communicate without available agents. Our system, as described, is intended to provide a platform-agnostic, flexible, and programmable path-calculation mechanism that, operating at the application layer, enables the scalability and future-proofing of modern routing solutions as well as the integration with future network architectures such as SONs in the Sixth-Generation (6G) networks.

        \begin{figure*}[ht!]
            \centering\includegraphics[keepaspectratio=true,scale=0.06]{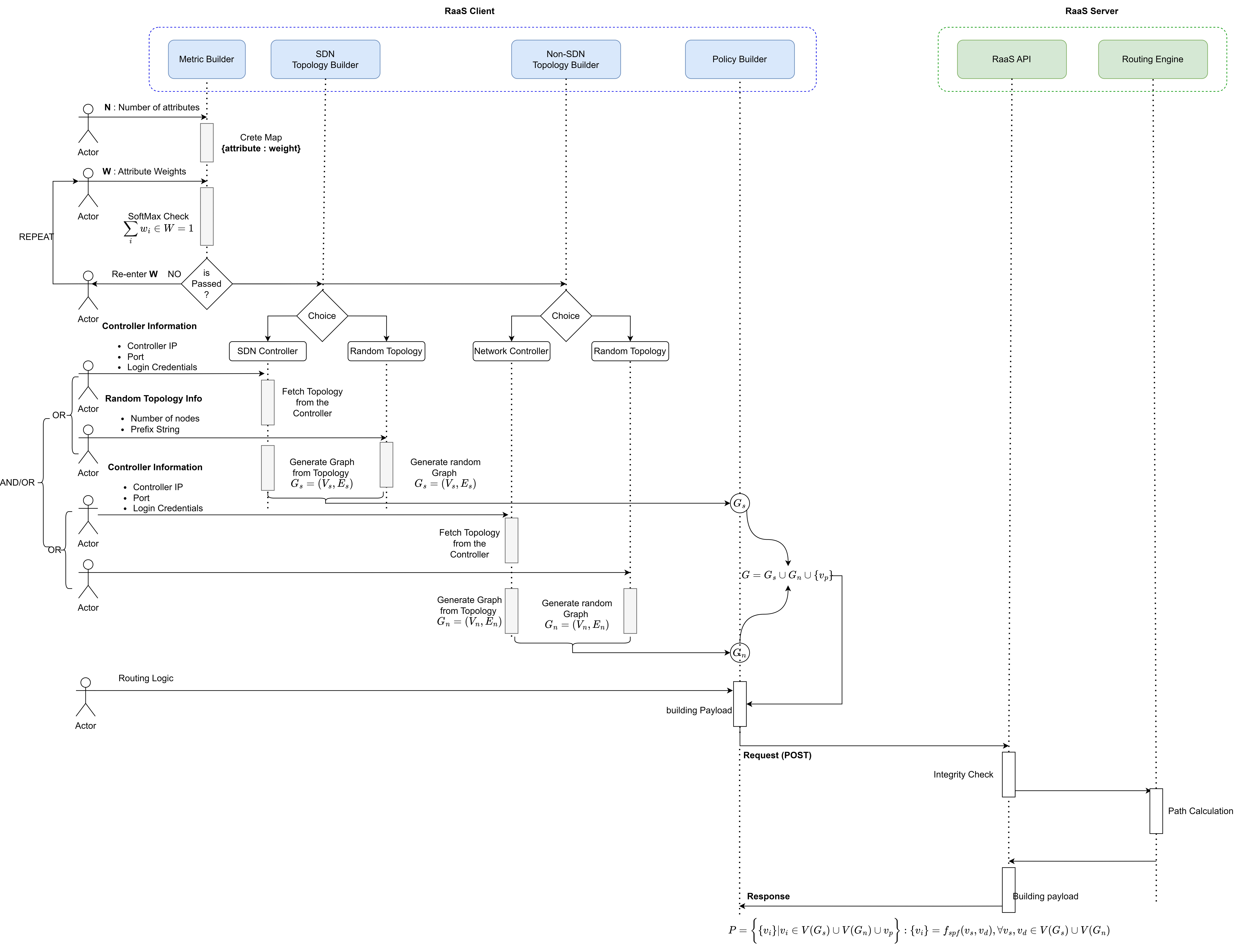}
            \caption{Sequence diagram of RaaS Client-Server model for a Hybrid-SDN topology}
            \label{fig:raas_seq}
        \end{figure*}

        \subsection{Routing-as-a-Service}
        RaaS is a data plane-agnostic principle of flexible and programmable path calculation mechanism served at the application layer~\cite{BU2019130, Lakshminarayanan:CSD-04-1327}. In our system, we define the Routing Logic as a pair of metric functions $f_{metric}$ and shortest-path algorithm $f_{sp}$. Where, $f_{metric}(w_ia_i|i\in[1,n], i\in \mathbb{N})$ is an arbitrary multivariate scalar function with a finite dependent variable set $A=\{a_i\}$, called attributes, weighed by corresponding weighing factor $w_i\in W$. $f_{sp}$ takes a weighted simple graph (i.e., free of self-loops and parallel-edges) $G(V,E)$, where $V$ and $E$ are the vertices (nodes) and edge sets of $G$ respectively, with a pair of nodes $v_s, v_d \in V$ and returns optimal path(s) $P_{s,d}\in 2^E$ such that path cost is minimal. Therefore, considering an Intent-Based Networking (IBN) paradigm, a RaaS application must accept the routing logic $(f_{metric}(A,W), f_{sp}(G(V,E)))$ as an \textit{intent} though an open-API, and returns the optimal path(s) $\{P_{s,d}\}$ as the response.    

        Figure \ref{fig:raas_seq} depicts the sequence diagram of a RaaS application operating in Client-Server mode on a Hybrid-SDN topology. As described in Section~\ref{sec:architecture}, splitting the application into client and server side results in load balancing and higher scalability. Where the client-side operations are I/O-intensive due to intent-API exposure and network controllers interfacing, the server-side is more compute-intensive as it runs the graph algorithms. The operation pipeline comprises the following phases:
        \begin{enumerate}
            \item \textit{Metric Building}: In this phase, the attribute set $A$ such that $|A|=N$ is provided with their corresponding weights $W$ along with a metric function $f_{metric}$. The weighing factor $w_i\in W$ of an attribute $a_i\in A$ signifies the priority of the attribute in the $f_{metric}$ definition. therefore, the sum of all $w_i$ must be $1$. 

            \item \textit{Topology Building}: In this phase, the iRaaS client interfaces with the network controllers to fetch their underlying topology. A network controller abstracts the platform-level information of the network and returns a graph representation of the topology. In a hybrid-SDN scenario, the RaaS client interfaces with each instance of the SDN and Non-SDN controllers to fetch their underplaying topologies into a set of graphs $\mathbb{G}=\{G_j(V_j,E_j)\}$. In such case, we propose a normalisation method of transforming $\mathbb{G}$ into a fused-graph $G(V,E)$ by adding a pseudo-node $v_p$ which is fully adjacent with a designated node $v_{dn}\in V_j$ from each graph $G_j$. An ideal designated node is preferably but not necessarily to be at the centre of its topology. This mimics the behaviour of a summary point in traditional routing protocols. The cost of the logical link between each $v_{dn}$ and $v_p$ is an explicitly defined non-zero value. This is because the routing between topologies across controllers under the same administrative domain requires the inter-controller data path to be involved, which is considered as an exterior link. $v_p$ represents the exterior link, and the cost between $v_p$ and $v_{dn}$ represents the cost of accessing it. 

            \item \textit{Policy Building}: The iRaaS Client in this phase prepares the routing logic by packing the normalised topology $G(V,E)$ and the metric function and sending it to the iRaaS Server for processing. 

            \item \textit{Integrity Check}: The iRaaS server checks the integrity of the routing logic received from the iRaaS client and passes it to the routing engine upon successful validation.

            \item \textit{Path Calculation}: The iRaaS servers calculates link-costs by applying the $f_{metric}$ function on telemetry data and returns the optimal path(s) as per the shortest path algorithm. 
        \end{enumerate}
        
       \subsection{Intelligent Routing Behaviours}
       The iRaaS Client and Server components provide the underlying framework for making intelligent routing decisions. In this section, we describe the set of behaviours that are essential for an intelligent routing model. In either case of using a heuristic algorithm or an ML model, these behaviours are critical for providing an optimal routing path across different domains. Moreover, for the graphs aggregated in our application plane, two algorithms are described that can minimise the footprint of our application and provide a more robust implementation.

        \subsubsection{Topology Aggregation}
        The aforementioned developments rely on typologies either reactively or proactively and maintain a local database for it. An iRaaS application must leverage the network controllers' north-bound interface to fetch and aggregate individual topologies into a global topology map. The aggregation process could be arbitrary, however, the graph aggregation algorithm must consider the inter-controller connectivity cost while taking the union of the candidate graphs. In the previous subsection, we took a simplistic approach by introducing a logical pivot-vertex $v_p$ per aggregation and a designated vertex $v_{dn}$ per candidate graph with a non-zero constant cost between them representing the inter-controller communication cost.   

        \subsubsection{Cost Normalization}
        The aggregated graph generated from the Topology Aggregation phase is an in-memory data structure at the iRaaS application server. The telemetry system monitors and maintains a database of network key performance indicators (KPIs); the metric function uses a subset of selected attributes from the KPIs to weight the aggregated graph. Traditional routing protocols only consider link costs for path-finding. However, in a softwarised network infrastructure where network functions are not necessarily physical, computational costs are also significant in calculating end-to-end costs. Therefore, a composite metric that considers both link and computational cost into a normalised cost is required, resulting in a more accurate routing decision. That said, normalising computational cost with link cost requires an isomorphic transformation of the aggregated graph. Since the computation load appears as a weighted self-loop, hence, it makes the graph a regular graph rather than a simple one. A simple graph is a prerequisite for running any shortest-path algorithm on it. Authors in \cite{sten} present such a normalization technique named Stochastic Temporal Edge Normalisation (STEN). 

        \subsubsection{Redundant Path Discovery}
        To ensure rapid convergence, the path-finding algorithm $f_{sp}$ must not only discover the best path but a set of alternate or redundant paths. This may resemble the Feasible Successor approach of DUAL which allows rapid path switchover in case of a primary path failure without involving any diffusion updates in the topology. However, DUAL's loop-prevention mechanism may not find any  Feasible Successor despite having available alternate paths if none of the non-successor neighbours pass the Feasibility Criteria. In such a case, EIGRP puts DUAL into an active state and initiates querying neighbours for alternate path discovery. This issue occurs because EIGRP is a distance vector routing protocol. In RaaS, the global topology-building process follows the link-state approach. Therefore with a global view of the topology, an explicit loop-prevention mechanism is unnecessary and hence, $f_{sp}$ can safely discover and maintain alternate paths to reactively switch between them when the primary one fails.

        \subsubsection{Reliability-Based Metric and Reactive Route Ranking}
        As the KPIs fluctuate over time with the network dynamics, it may invoke $f_{sp}$ unnecessarily many times, resulting in inefficient runtime behaviour such as route-flapping and high computational complexity. We propose two approaches to optimise the runtime footprint for RaaS applications. 

        The first one is Reactive Route Ranking where $f_{sp}$ processes the aggregated graph in two phases. In Phase-1, it calculates all possible paths between all pairs of vertices. Defining a cut-off diameter and initializing the link costs with seed values speed up the process further. As the path discovery between each pair is sequentially independent, therefore they can run simultaneously. The result is a forest of Shortest Path Trees (SPT) where each instance is rooted by the destination vertex $v_d$ with all branches representing a unique path to the source $v_s$ as the identical leaf for all branches. In Phase-2, a Raas application may calculate the path cost of each branch for all SPTs and rank the paths of each SPT. The two-phase approach limits re-convergence to only topology change scenarios. If the topology grows then only new connections are updated in the SPTs, and if it shrinks, then vanishing links are updated with an infinite cost.

        Second, we propose the use of Reliability as a metric. Reliability is statistically calculated using Sharpe Ratio \cite{sharpe} from a rolling window of $f_{metric}$. RaaS application uses the expected Reliability from a Recurrent Neural Network (RNN) to make routing decisions. Therefore, iRaaS avoids fluctuating unreliable routes and emphasizes more on reliable paths rather than the shortest, least-costly and fastest paths. 

        In summary, the above two optimization steps ensure minimum invocation of re-convergence with Reactive Route Ranking and prioritize reliable routes. Authors in \cite{kdn} have explained the above techniques in detail.

    \subsection{Telemetry Framework and its Communication Modes}
    \begin{figure*}[ht]
        \centering
        \includegraphics[keepaspectratio=true,scale=0.9]{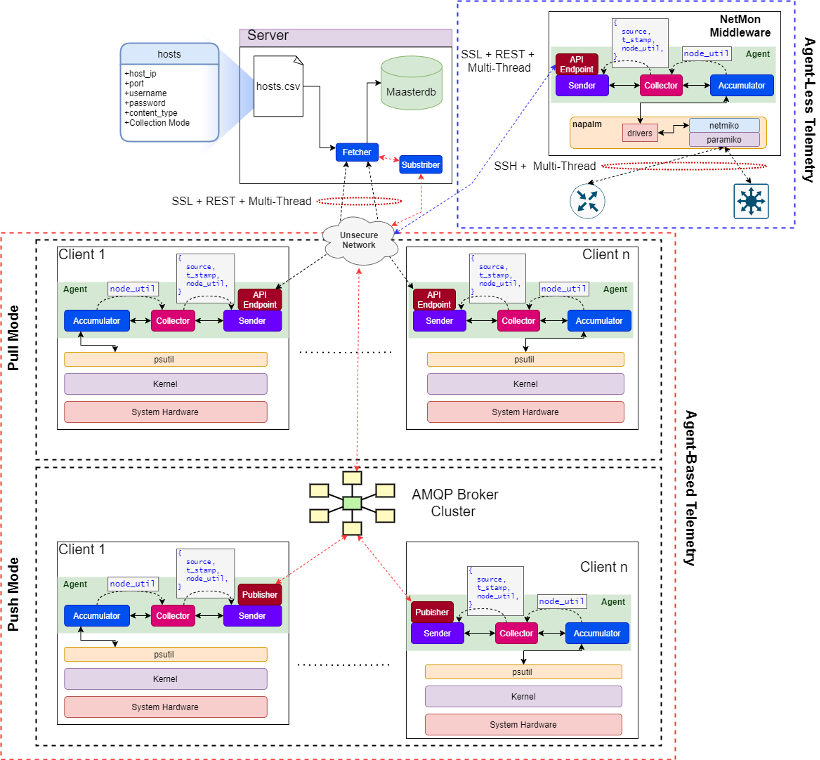}
        \caption{System Level Diagram of the Telemetry System}
        \label{fig:telemetry}
    \end{figure*}

        This section describes the telemetry framework architecture, which carries the monitoring data from the data to the application plane. The architecture of this system can be seen in Figure \ref{fig:telemetry}. We propose a modular, multi-modal telemetry API named \textit{ShellMon} for the RaaS framework. Unlike vendor, platform and version-dependent telemetry protocols such as SNMP \cite{rfc5343}, NetFlow \cite{rfc3954}, \textit{ShellMon} provides platform and vendor agnostic multi-modal telemetry with a common standard data format. Table \ref{tab:ShellMon-specs} summarises the operational specifications of \textit{ShellMon}. 

        \begin{table}[t]
            \caption{Operational specification of \textit{ShellMon}}
            \resizebox{\columnwidth}{!}{%
            \begin{tabular}{|l|l|l|ll}
            \cline{1-3}
            Attributes                          & Type                                                                         & Usage                                       &  &  \\ \cline{1-3}
            \multirow{2}{*}{Mode of Transport}  & \begin{tabular}[c]{@{}l@{}}Mode 1: Request-Response\\ (RESTful)\end{tabular} & Pull based telemetry collection             &  &  \\ \cline{2-3}
                                                & \begin{tabular}[c]{@{}l@{}}Mode 2: Publish-Subscribe \\ (AMQP)\end{tabular}  & Push based telemetry collection             &  &  \\ \cline{1-3}
            \multirow{2}{*}{Mode of Collection} & Mode 1: Agent Based                                                          & Target device with monolithic karnel        &  &  \\ \cline{2-3}
                                                & Mode 2: Agent Less                                                           & Target device allows agent installation     &  &  \\ \cline{1-3}
            Serialisation Format                & JSON                                                                         & Standard data format, low parsing footprint &  &  \\ \cline{1-3}
            \end{tabular}%
            }
            \label{tab:ShellMon-specs}
        \end{table}

         \textit{ShellMon} server maintains a \textit{host} file containing the clients' information (e.g., hostname, port number, access credentials, the content type of the payload and connection mode). The \textit{Fetcher} and \textit{subscriber} modules use request-response and publish-subscribe modes, respectively. The telemetry is collected and stored in a Master database shared with the iRaaS application. The remainder of this section describes the various modes of communication that \textit{ShellMon} offers.

        \subsubsection{Transport Mode 1: RESTfull Request-Response}
        In this mode, the collection mechanism operates in a RESTful fashion. \textit{ShellMon} server sends poll requests in regular intervals, which triggers the clients to invoke device-level local KPI collection. Clients timestamp the KPI samples and send them back to the server. That said, this mode relies on HTTP’s keep-alive mechanism to monitor the liveliness of the clients. 

        \subsubsection{Transport Mode 2: Publish-Subscribe }
        This mode is suitable for large-scale client-base, where the number of ports available on the server side is constrained. The agent comprises identical modules as of the RESTful mode. However, instead of an API end-point, it publishes KPIs from a local publisher. \textit{ShellMon} uses an Advanced Message Queuing Protocol (AMQP) \cite{amqp} message broker for the transport. 
                
        \subsubsection{Collection Mode 1: Agent-Less Telemetry}
            This mode is suitable for network devices running monolithic kernels such as Cisco IOS and Juniper JunOS, which do not allow the installation of external agents. The \textit{NetMon} middleware communicates with the network devices through asynchronous SSH sessions to collect telemetry information using a multi-vendor SSH library called Napalm [ref-netmiko]. The Accumulator module of the NetMon agent collects monitoring data samples from Napalm into a key-value store named \textit{node\_util}. The Collector module tags it with \textit{source\_id} and timestamp. Finally, the Sender module exposes an API endpoint for polling the monitoring data. NetMon uses a request-response mechanism for polling, i.e., a poll request from the \textit{ShellMon} server initiates the collection cycle; therefore, the agent does not require any local queue. 
        
        \subsubsection{Collection Mode 2: Agent-Based Telemetry}
            An agent runs on top of the network device kernel in this mode. The Accumulator, Collector and Sender module behaves the same as the Agent-Less mode.

    \section{Test-bed Setup and Proof-of-Concept}
        \begin{figure}[t]
            \centering
            \includegraphics[keepaspectratio=true,scale=0.38]{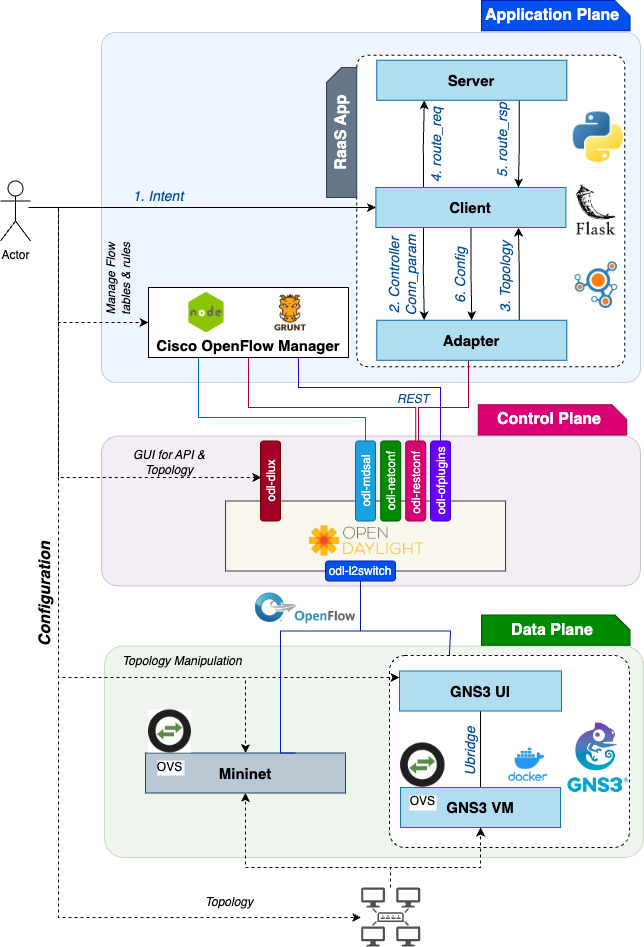}
            \caption{iRaaS Deployment Diagram}
            \label{fig:raas_dep}
        \end{figure}
        
        Figure \ref{fig:raas_dep} depicts the deployment diagram of the RaaS testbed along with the technology stack used. The application, control and data planes are segregated by three Virtual Machines (VMs) for runtime isolating. The data plane comprises a Mininet \cite{mn} and a GNS3 for simulating network topologies. Mininet supports Open-V-Switches (OVS) natively, we used a containerized version of OVS to deploy in GNS3. We use OpenDaylight (ODL) at the control plane to interface with OVS and the RaaS application using OpenFlow v1.3 and RESTCONF respectively. To carry out control plane operations, we use the following Opendaylight features;  \textit{odl-l2swicth} module controls the south-bound interface using OpenFlow, \textit{odl-restconf} controls the north-bound interface using RESTCONF, \textit{odl-mdsal} provides Model-Driven Service Abstraction Layer to parse YANG data-models, \textit{odl-ofplugin} provides a standard interface between the control and data plane, and \textit{odl-dlux} provides ODL GUI. 

        The application plane hosts the iRaaS application and a Cisco OpenFlow Manager (OFM) VM. OFM inspects the topology from the ODL controller and provides a GUI-based flow management tool. The Client module of the iRaaS app receives intent from the administrator that includes the controller access information and routing logic. The Client interfaces with the Adapter module to fetch topology from the data plane following the methods specified in figure \ref{fig:raas_seq}. Further, it sends a \textit{route\_request} to the Server Module which computes the optimal paths and replies with a \textit{route\_response}. In this setup, we have developed the iRaaS app using Flask micro-framework for API development and the \textit{NetworkX} library for graph computation. 

\begin{figure}
            \centering
            \includegraphics[width=0.9\linewidth]{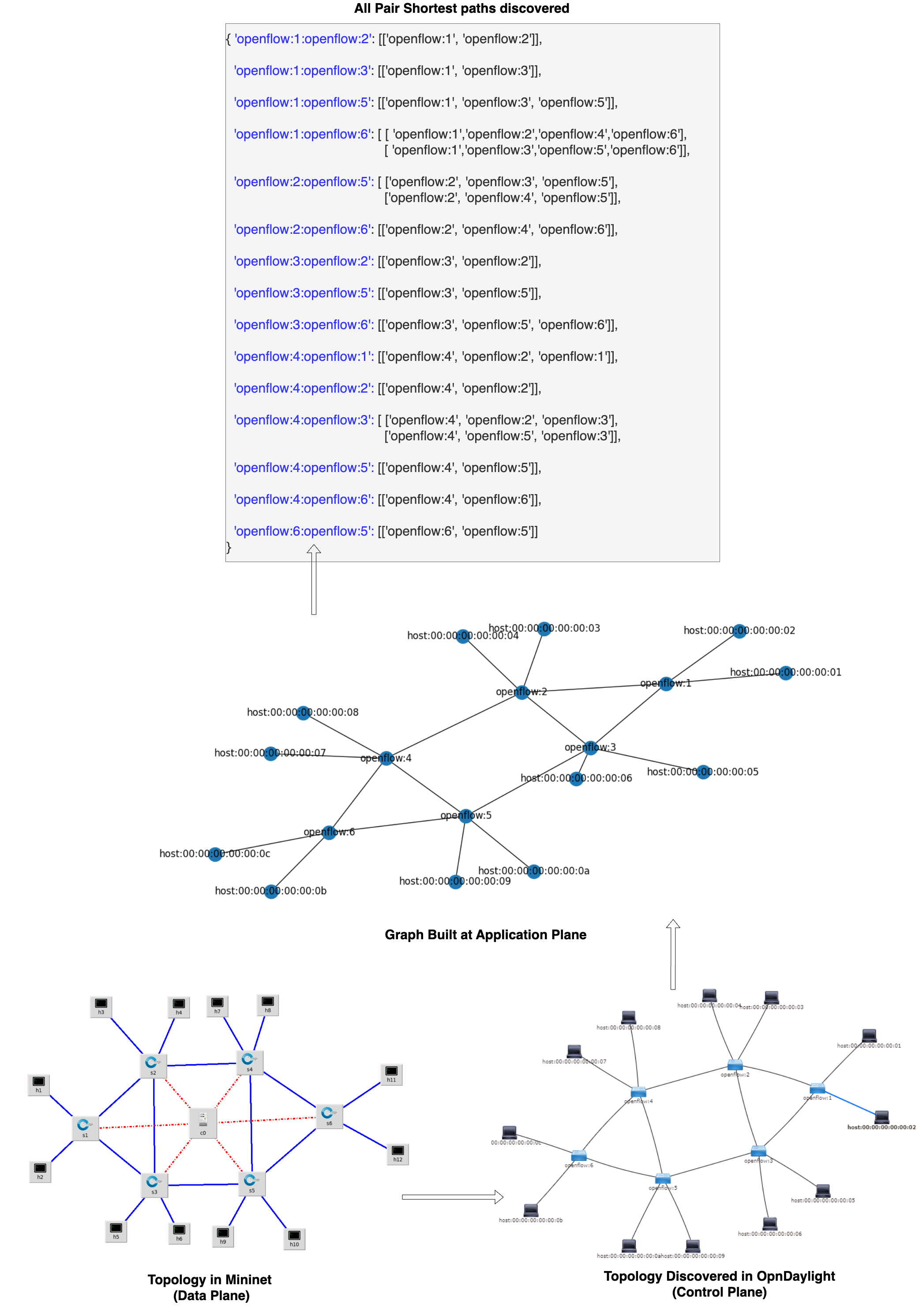}
            \caption{iRaaS calculates All-pair shortest path at the application plane}
            \label{fig:poc_result}
        \end{figure}

        Figure \ref{fig:poc_result} depicts the flow of topology processing from the Mininet data plane to the RaaS application through the OpenDaylight control plane. The shown example illustrates a partial mesh topology of six Open-V-Switch instances each connecting two hosts and communicating with a remote SDN controller over OpenFlow v1.3. iRaaS adapter fetches the topology from the OpenDaylight controller and the iRaaS client builds a graph data structure using the NetworkX library as shown in the figure. In this test, we choose all-pair Dijkstra's algorithm to find all routes between each pair of OpenFlow switches as shown at the top of the figure. 
        
        The above test results validate the proof of concept of the proposed architecture. However, the same is also capable of running customized routing algorithms by altering configuration at the iRaaS server. That said, the paper \cite{kdn} shows rapid convergence in Knowledge-Defined Networks comparing the scalability against SPF and DUAL. 
    \newpage
    \section{Conclusion and Future Scope}
    This paper presents a system-level architecture of Intelligent Routing as a Service (iRaaS), a sequence diagram explaining the data between various iRaaS components and a robust telemetry architecture for collecting monitoring data from the underlying network infrastructure. A proof of concept setup also validates the operational capabilities of the proposed architecture with a deployment diagram detailing the assembly of various open-source components constituting the test bed used for experiments. 

    We aim to advance the iRaaS concept with a robust cognitive plane comprising additional machine learning-based operations algorithms such as traffic classification and state prediction.

    \section*{Acknowledgement}
    This work is a contribution by Project REASON, a UK Government funded project under the Future Open Networks Research Challenge (FONRC) sponsored by the Department of Science Innovation and Technology (DSIT).
    
    \bibliographystyle{IEEEtran}
    \bibliography{main}

\end{document}